# A Casimir approach for radiative self-energy

## Allan Rosencwaig*

### *Arist Instuments, Inc*
### *Fremont, CA 94538*

## Abstract

We apply a Casimir energy approach to evaluate the self-energy or one-photon radiative correction for an electron in a hydrogen orbital. This linking of the Lamb shift to the Casimir effect is obtained by treating the hydrogen orbital as a one-electron shell and including the probability of the electron being at a particular radius in that orbital and the probability that the electron will interact with a virtual photon of a given energy.

* E-mail address: allan@aristinst.com

The Casimir effect [1] and the Lamb shift [2] are considered to be two compelling observations that indicate the reality of vacuum fluctuations of the quantum electromagnetic field. The Casimir effect arises from the influence of boundaries on the field and the Lamb shift is the result of coupling between an atomic electron and the quantum field through the emission and absorption of virtual photons. Although both effects are due to the presence of vacuum fluctuations of the quantum electromagnetic field, they are generally considered to be separate and distinct. However, we will show that on the atomic scale  the two effects are closely related and that, in particular, the atomic Casimir effect can account for the principal component of the observed Lamb shift, the one-photon radiative correction or self-energy.

The Lamb shift, discovered by Lamb and Retherford in 1947, is a small energy splitting in the $2s_{1/2}$ and $2p_{1/2}$ levels of hydrogen, levels which are fully degenerate even when treated relativistically. The observed Lamb shift of 1057.58 MHz is actually the net result of several different energy shifts arising from different contributions; self-energy or one-photon radiative correction, vacuum polarization, relativistic recoil, two- and three-photon radiative corrections and higher order contributions [3]. All of these contributions have been very accurately calculated through QED, and the extremely close agreement between theory and experiment is one of the well-known successes of QED.

The most important of the terms that make up the Lamb shift is the self-energy or one-photon radiative correction of the atomic electron that results from the emission and absorption of a virtual photon from the vacuum electromagnetic field. The general form of the self-energy is given by [4],

$$E_{SE} = \frac{1}{n^3}\alpha^5 mc^2 F(n,l) \qquad\qquad (1)$$

Where $n$ and $l$ are the principal and orbital angular momentum quantum numbers of the bound electron, $\alpha$ is the fine-structure constant (1/137), $m$ is the electron mass and $c$ is the velocity of light. The term $F(n,l)$ is a numerical factor that depends on both quantum numbers. In general, $F(n,l)$ is of order 3 for s states ($l = 0$), and of order $\pm 0.04$ for non-s states ($l > 0$). For any given principal quantum number, $n$, the self-energy for s-states is much greater than for non-s states and while the self-energy term is always positive, i.e. repulsive, for s- states, it can be either repulsive or attractive for non-s-states.

The Casimir energy was originally calculated for the macroscopic case of two perfectly conducting parallel plates of area $L^2$ that are separated by a distance $d$ where $d << L$. The Casimir energy arises from the fact that the normal modes of the quantum electromagnetic field between the two plates must be zero at the plate boundaries. This then limits the modes that can exist within the gap $d$, with the result that there are more modes outside the gap than inside. Thus the radiation pressure is greater outside and the Casimir energy is found to be [5],

$$U_p = -\frac{\pi^2 \hbar c}{720 d^3} L^2 \tag{2}$$

where $\hbar$ is the reduced Planck's constant.

The Casimir energy for the 2-plate configuration is, as expected, negative and hence attractive. In fact, the Casimir energy is attractive for most boundary configurations. However, for the very important case of a perfectly conducting hollow sphere or shell of radius $r$, it is found to be repulsive and is given by [6-8],

$$U_s = 0.0462 \frac{\hbar c}{r} \tag{3}$$

The change from attractive to repulsive in going from parallel plates to a sphere is generally attributed to a change in the boundary conditions for the electromagnetic field, but there does not appear to be any simple explanation for the repulsive character of the Casimir energy within a perfectly conducting sphere.

The basic concepts of the Casimir theory have been extended to the microscopic region allowing atomic and molecular effects, such as the van der Waals force, to be understood as a variation on the theme of the Casimir effect [9,10]. Casimir energy concepts are also used in the chiral bag model of the nucleon [11].

We will attempt to calculate, from a Casimir perspective, the self-energy for an electron in a hydrogenic state by treating the atomic orbital as if it were a one-electron shell. This is clearly not a bulk perfectly conducting shell. Thus we must include the probability of the electron being at a radius $r$ in the orbital, and the probability that the electron will interact with a virtual photon of a given energy. We accomplish the former by calculating electron probabilities through the hydrogen wavefunction, and we account for the latter by including the electromagnetic scattering cross-section. Combining these two probabilities as $P_s(r)$, we have,

$$P_s(r) = \frac{\sigma_T P_{nlm}(r)}{kr^2} \tag{4}$$

where $\sigma_T$ is the total scattering cross-section and $P_{nlm}(r)$ represents the probability of the electron in an $(nlm)$ orbital being at the radial position $r$. The denominator, $kr^2$ where $k$ is a number, normalizes the scattering cross-section to the radius of the shell, such that the probability of the electron interacting with the virtual photon $\rightarrow 1$ as $\sigma_T \rightarrow r^2$. We have obtained our best results at $k \approx 1$.

Let us first consider the case where we set $\sigma_T$ to the constant Thomson scattering cross-section,

$$\sigma_T = \frac{8\pi}{3} a_0^2 \qquad (5)$$

where $a_o$ is the classical electron radius $e^2/mc^2$. This is equivalent to treating the problem in a non-relativistic fashion. In the following treatment we shall also make use of the following important relationship,

$$a_0 = \alpha \hbar_c = \alpha^2 a \qquad (6)$$

where $\hbar_c = \hbar/mc$ is the reduced Compton wavelength of the electron, and $a$ is the atomic Bohr radius.

For a constant $\sigma_T$, we can modify Eqn. (3) to obtain the Casimir energy at a radius $r$ in an $(nlm)$ orbital,

$$U_C(r) = \frac{8\pi}{3} (0.0462) \frac{\hbar c}{r} \frac{a_0^2}{r^2} P_{nlm}(r) \qquad (7)$$

For orbitals such as the $2s_{1/2}$ and $2p_{1/2}$ states that are given by the basic hydrogen eigenstates,

$$P_{nlm}(r) = \int |\psi_{nlm}(r,\theta,\varphi)|^2 \sin\theta d\theta d\varphi \qquad (8)$$

where $\psi_{nlm}$ is the hydrogen wavefunction for the $(nlm)$ state and $\theta$ and $\varphi$ are the polar and azimuthal angles respectively of the spherical coordinate system. The general expression for the $(nlm)$ hydrogen eigenstate is given by,

$$\psi_{nlm} = \sqrt{\left(\frac{2}{na}\right)^3 \frac{(n-l-1)!}{2n[(n+l)!]^3}} e^{-r/na} \left(\frac{2r}{na}\right)^l L_{n-l-1}^{2l+1}\left(\frac{2r}{na}\right) Y_l^m(\theta,\varphi) \qquad (9)$$

where $L$ is the Laguerre polynomial and $Y$ is the spherical harmonic. Since $\int |Y_l^m(\theta,\varphi)|^2 \sin\theta d\theta d\varphi = 1$, we can write $P_{nlm}$ as,

$$P_{nlm}(r) = \Phi_{nl}^2 \left( r/a \right) = \frac{1}{n^3 a^3} \xi_{nl}^2 \left( r/a \right) \qquad (10)$$

where $\Phi_{nl}$ is the usual radial part of the hydrogen eigenfunction .

Thus the Casimir energy integrated over the (nlm) orbital becomes,

$$U_C = \int U_C(r) r^2 dr = \frac{8\pi}{3}(0.0462)\hbar c \left\{ \frac{1}{n^3} \frac{a_0^2}{a^3} \int_{r_{min}}^{\infty} \xi_{nl}^2 \left( r/a \right) \frac{dr}{r} \right\} \qquad (11)$$

But,

$$\hbar c \frac{a_0^2}{a^3} = \alpha^5 mc^2$$

Therefore, the Casimir energy integrated over the (nlm) orbital becomes,

$$U_C(n,l) = \frac{1}{n^3} \alpha^5 mc^2 K(n,l) \qquad (12)$$

with

$$K(n,l) = \frac{8\pi}{3}(0.0462) \int_{r_{min}}^{\infty} \xi_{nl}^2 \left( r/a \right) \frac{dr}{r} \qquad (13)$$

We can see that the Casimir energy in Eqn. (12) has the same form as the one-photon radiative correction or self-energy of Eqn. (1). The lower bound for the integral in the expression for $K$ is a minimum radius that corresponds to our limitation on the highest photon energy to $mc^2$, and thus the shortest photon wavelength to $\lambda_c$. For a spherical shell, most of the electromagnetic modes are spherical modes that propagate around the circumference, the "whispering-gallery modes". However, the largest $\lambda$ at any $r$ is not given by a circumferential mode but by $\lambda = 2r$. This then sets $r_{min}$ at ½ $\lambda_c$.

In Table I we list the $\xi_{nl}^2/n^3$ ($= a^3\Phi_{nl}^2$) expressions for different hydrogen atomic levels. Using the expression for the $2s_{1/2}$ level in Eqn. (14), we find that the Casimir energy in frequency for this level is 1233 MHz as compared to the QED value [3] of the one-photon radiative correction or self-energy of 1069 MHz. This is a reasonably close agreement considering the non-relativistic assumption of a constant scattering cross-section.

If we include relativistic terms, the total electromagnetic scattering cross-section is a decreasing function of the photon energy. The relativistic differential scattering cross-section is given by the Klein-Nishina formula,

$$\frac{d\sigma}{d\Omega} = \frac{1}{2}a_0^2 \left\{ \frac{1}{[1 + \varepsilon(1 - \cos\vartheta)]} - \frac{\sin^2\vartheta}{[1 + \varepsilon(1 - \cos\vartheta)]^2} + \frac{1}{[1 + \varepsilon(1 - \cos\vartheta)]^3} \right\}$$

$$\varepsilon = \frac{\hbar\omega}{mc^2}$$

(14)

where $\hbar\omega$ is the vacuum photon energy and $\vartheta$ is the scattering angle. For low photon energies, the differential scattering cross-section takes on the classical form,

$$\frac{d\sigma}{d\Omega} = \frac{1}{2}a_0^2 \left(1 + \cos^2\vartheta\right)$$

(15)

The total scattering cross-section is derived by integrating over the solid angle, and for the relativistic Klein-Nishina formula we have,

$$\sigma_T = 2\pi a_0^2 f(\varepsilon)$$

(16)

with

$$f(\varepsilon) = \frac{(2 + 6\varepsilon + \varepsilon^2)(1 + \varepsilon)^2 - (2 + 6\varepsilon + 5\varepsilon^2) + 2(1 + \varepsilon)^2(-2 - 2\varepsilon + \varepsilon^2)\ln(1 + \varepsilon)}{2\varepsilon^3(1 + \varepsilon)^2}$$

(17)

For $\varepsilon << 1$ it is straightforward to show that $f(\varepsilon) = 4/3$ and $\sigma_T = \frac{8\pi}{3}a_0^2$.

It is clear that a total scattering cross-section that decreases with photon energy will decrease the Casimir energy. This is analogous to the decrease in Casimir energy for a macroscopic conducting shell when the photon frequency exceeds the plasma frequency [12]. To calculate the effect of an energy-dependent cross-section, we need to apply the function $f(\varepsilon)$ to the initial calculation of the Casimir energy of Eqn. (3). This is difficult to do since the derivation of Eqn. (3) is quite complicated. However, a derivation by Klich [13], who uses a contour integral method for the mode summation in a spherical shell, appears amenable to such a modification.

From Klich, the Casimir energy for a perfectly conducting shell of radius $r$ can be written as,

$$U_C = \frac{\hbar c}{4\pi} \int_0^\infty G(r, z)\, dz$$

(18)

where z = $1/\lambda$ and,

$$G(r, z) = e^{-4rz}\left(1 + 4rz + 4r^2z^2\right)$$

(19)

Setting $y = r/\lambda_c$ and since $\varepsilon = \lambda_c/\lambda$, the function $G$ can be written as,

$$G(r,z) = G(y,\varepsilon) = e^{-4y\varepsilon}\left(1 + 4y\varepsilon + 4y^2\varepsilon^2\right) \tag{20}$$

Incorporating the scattering cross-section into Eqn. (18), the Casimir energy for a shell of radius $r$ that has only one electron in that shell would then be,

$$U_C = \frac{\hbar c}{4\pi r^2}\int_0^\infty \sigma_T(z)G(r,z)\,dz = \frac{1}{2}a_0^2\frac{\hbar c}{\lambda_c r^2}\int_{\varepsilon_{\min}}^\infty f(\varepsilon)G(y,\varepsilon)\,d\varepsilon \tag{21}$$

where,

$$\varepsilon_{\min} = \frac{\lambda_c}{\lambda_{\max}} = \frac{\lambda_c}{2r} \tag{22}$$

The analytical integral in Eqn. (21) is quite difficult, but we can simplify $f(\varepsilon)$. If we maintain an upper bound to the photon energies at $mc^2$, we find that a good fit to $f(\varepsilon)$ from $\varepsilon = 0$ to $1$ is,

$$f_1(\varepsilon) \approx 1.331 - 1.0723\varepsilon + 0.84431\varepsilon^2 - 0.30862\varepsilon^3 \tag{23}$$

Using $f_1(\varepsilon)$ in the integral of Eqn. (21) and limiting the integration to the range from 0 to 1 we obtain the new integral,

$$I(y) = \int_0^1 f_1(\varepsilon)G(y,\varepsilon)\,d\varepsilon$$

$$= \left\{\frac{0.832\left(-0.380+y\right)\left(0.229+0.018y+y^2\right)}{y^4} - \frac{0.794e^{-4y}\left(0.527+y\right)\left(0.225+0.265y+y^2\right)\left(0.769+1.665y+y^2\right)}{y^4}\right\} \tag{24}$$

$$= g(y)$$

For a hydrogen orbital we need to include the probability $P_{nlm}(r)$. We then have for the Casimir energy,

$$U_C(nlm) = \frac{1}{2n^3}\hbar c\frac{a_0^2}{a^3}\frac{1}{\lambda_c}\int_{r_{\min}}^\infty g(y)\xi_{nl}^2\left(r/a\right)dr \tag{25}$$

Setting $x = r/a = 2\pi\alpha r/\lambda_c$ and $y = r/\lambda_c = x/2\pi\alpha$, we then have,

$$U_C\left(nlm\right) = \frac{1}{n^3}\alpha^5 mc^2 \left\{ \frac{1}{4\pi\alpha} \int\limits_{x_{\min}}^{\infty} g\left(x\right)\xi_{nl}^2\left(x\right)dx \right\} \qquad (26)$$

where the lower bound $x_{min} = \lambda_c/2a = \pi\alpha$. For the $2s_{1/2}$ level we now get $U_C = 1079$ MHz, which is within 1% of the QED value of 1069 MHz for the self-energy.

We can apply this model to non-s states as well provided that we account for two factors. First, since the non-s states are not spherical, they do not cover a full $4\pi$ of solid angle. We can take this into account by defining a spherical coverage factor $\beta = 1/\left(4\pi c_{nl}^2\right)$ where $c_{nl}$ is the normalizing pre-factor of the corresponding spherical harmonic $Y_l^m$. Secondly, we know that for boundary configurations far from spherical, the Casimir energy becomes attractive rather than repulsive. We can take both factors into account by including in Eqn. (26) the term,

$$\eta = 2\left(\beta - \frac{1}{2}\right) \qquad (27)$$

The Casimir energy for any (nlm) hydrogen level then becomes,

$$U_C\left(nlm\right) = \frac{1}{n^3}\alpha^5 mc^2 \left\{ \frac{\eta}{4\pi\alpha} \int\limits_{x_{\min}}^{\infty} g\left(x\right)\xi_{nl}^2\left(x\right)dx \right\} \qquad (28)$$

For all s- states, $\beta = 1$ and $\eta = 1$. For the $2p_{1/2}$ state, $\beta = 1/3$ and $\eta = -1/3$, while for the $2p_{3/2}$ state, which is actually an admixture of (211) and (21-1) states, $\beta = 2/3$ and $\eta = 1/3$. For the $2p_{1/2}$ state the Casimir energy becomes -14.5 MHz while the QED value for the self-energy is -12.8 MHz [3]. In Table II we list the Casimir energies obtained from Eqn.(28) and the QED self-energies for a number of hydrogen atomic levels. We see very good agreement for all s levels and quite good agreement for the 2p1/2 and 2p3/2 levels.

Although a full QED calculation is needed to obtain the most accurate values for the self-energy or radiative correction for electrons in hydrogenic orbits, it is interesting to see that quite good estimates of this self-energy can be obtained by this much simpler Casimir energy approach. This ability to use a Casimir approach to obtain self-energies can be applied to other areas as well.

## Acknowledgements

The author would like to thank L. Kofman and J.R. Bond for their insightful comments.

## **<u>TABLE CAPTIONS</u>**

Table I:  The radial function $\zeta^2_{nl}/n^3$ ($= a^3\Phi^2_{nl}$) for a number of different hydrogen orbitals (*nl*).

Table II: The parameters $\beta$ and $\eta$ and the one-photon radiative correction or self-energy, in MHz, obtained from the Casimir approach and through QED for different hydrogen orbital states.

**<u>Table I</u>**

| **(nl)** | $\dfrac{1}{n^3}\xi_{nl}^2 = a^3\Phi_{nl}^2$ |
|---|---|
| (10) | $4e^{-2r/a}$ |
| (20) | $\dfrac{1}{2}\left(1-\dfrac{1}{2}\dfrac{r}{a}\right)^2 e^{-r/a}$ |
| (21) | $\dfrac{1}{24}\left(\dfrac{r}{a}\right)^2 e^{-r/a}$ |
| (30) | $\dfrac{4}{27}\left[1-\dfrac{2}{3}\dfrac{r}{a}+\dfrac{2}{27}\left(\dfrac{r}{a}\right)^2\right]^2 e^{-2r/3a}$ |
| (40) | $\dfrac{1}{16}\left[1-\dfrac{3}{4}\dfrac{r}{a}+\dfrac{1}{8}\left(\dfrac{r}{a}\right)^2-\dfrac{1}{192}\left(\dfrac{r}{a}\right)^3\right]^2 e^{-r/2a}$ |

**<u>Table II</u>**

| orbital state | $\beta$ | $\eta$ | $\Delta\nu$(Casimir) (MHz) | $\Delta\nu$(QED) (MHz) |
|---|---|---|---|---|
| 1s | 1 | 1 | 8867 | 8367 |
| 2s | 1 | 1 | 1079 | 1069 |
| $2p_{1/2}$ | 1/3 | -1/3 | -14.5 | -12.8 |
| $2p_{3/2}$ | 2/3 | 1/3 | 14.5 | 12.8 |
| 3s | 1 | 1 | 318 | 326 |
| 4s | 1 | 1 | 134 | 137 |